%% file: conference_101719.tex
\documentclass[conference]{IEEEtran}
\IEEEoverridecommandlockouts
\usepackage[hyphens]{url}
\usepackage[hidelinks]{hyperref}
\usepackage[hyphenbreaks]{breakurl}
\usepackage{cite}
\usepackage{amsmath,amssymb,amsfonts}
\usepackage{algorithmic}
\usepackage{graphicx}
\usepackage{textcomp}
\usepackage{xcolor}
\usepackage[hidelinks]{hyperref}
\usepackage{orcidlink}
\usepackage{acronym}
\usepackage{cleveref}
\usepackage{subcaption}
\usepackage{enumitem}
\usepackage{comment}

\usepackage{tikz}
\usetikzlibrary{arrows.meta, positioning, calc, graphs, matrix, shapes, circuits.logic.CDH, fit}

\usepackage[]{todonotes}

\acrodef{PoW}{Proof of Work}
\acrodef{XRT}{Xilinx Run Time}
\acrodef{HBM}{High Bandwidth Memory}
\acrodef{RAMA}{Random Access Master Attachment}
\acrodef{PC}{Pseudo Channel}

\def\BibTeX{{\rm B\kern-.05em{\sc i\kern-.025em b}\kern-.08em
    T\kern-.1667em\lower.7ex\hbox{E}\kern-.125emX}}

\begin{document}
\input{figures/tikzconfig.tex}
\input{figures/AES_construction.tex}

\title{Mining CryptoNight-Haven on the Varium C1100 Blockchain Accelerator Card$^*$
\thanks{$^*$The authors gratefully acknowledge the support of Xilinx for the donation of a Varium C1100 through the Xilinx 2021 Adaptive Computing Challenge.}
}

\author{\IEEEauthorblockN{Lucas Bex\IEEEauthorrefmark{1}, Furkan Turan\orcidlink{0000-0002-0045-7794}\IEEEauthorrefmark{2}, Michiel Van Beirendonck\orcidlink{0000-0002-5131-8030}\IEEEauthorrefmark{2}, Ingrid Verbauwhede\orcidlink{0000-0002-0879-076X}\IEEEauthorrefmark{2}}
\IEEEauthorblockA{\textit{imec-COSIC KU Leuven}\\
Kasteelpark Arenberg 10 - bus 2452, 3001 Leuven, Belgium \\
\IEEEauthorrefmark{1} cas.bex@student.kuleuven.be \\
\IEEEauthorrefmark{2} \{firstname\}.\{lastname\}@esat.kuleuven.be\\}
}

\maketitle

\begin{abstract}

Cryptocurrency mining is an energy-intensive process that presents a prime candidate for hardware acceleration. This work-in-progress presents the first coprocessor design for the ASIC-resistant CryptoNight-Haven \ac{PoW} algorithm. We construct our hardware accelerator as a \ac{XRT} RTL kernel targeting the Xilinx Varium C1100 Blockchain Accelerator Card. The design employs deeply pipelined computation and \ac{HBM} for the underlying scratchpad data. We aim to compare our accelerator to existing CPU and GPU miners to show increased throughput and energy efficiency of its hash computations.

\end{abstract}


\section{Introduction}
\acresetall

Cryptocurrencies have become an increasingly popular medium for decentralized transactions in recent years. The computation of a cryptographic \ac{PoW} is the foundational concept of decentralized consensus. It is computed by so-called miners, which are rewarded with a fraction of cryptocurrency for their effort. Currently, \ac{PoW} computations are built by iterating a cryptographic hash function until the output has a dedicated form. This energy-intensive process presents a prime candidate for hardware acceleration.

Several cryptocurrencies adopt so-called ASIC-resistant \ac{PoW} algorithms. These types of \ac{PoW} aim to deter dedicated hardware miners in favor of CPU- and GPU-based miners, which are more generally available to the public, thereby sustaining the distributed ledger idea. The Haven Protocol \cite{Haven} is one of the projects with such a goal. For its \ac{PoW}, Haven leverages on a custom ASIC-resistant hash function named as \textit{CryptoNight-Haven}.

In this project, we challenge the ASIC-resistance claims of CryptoNight-Haven, by implementing the \ac{PoW} as an RTL kernel on FPGA. We target the Xilinx Varium C1100 Blockchain Accelerator Card \cite{VariumC1100}. The card employs Xilinx' recent Ultrascale+ architecture, \ac{XRT} integration over a high-speed PCIe Gen 4 bus connection, and 8~GB of \ac{HBM}. Xilinx demonstrated that the Varium C1100 accelerates transaction validation in Hyperledger Fabric \cite{androulaki2018hyperledger} over an Intel Xeon Silver 4114 CPU with a factor of 14$\times$.

Our CryptoNight-Haven accelerator aims at a full hardware-based computation of the hash rather than software-assisted. It employs a pipelined datapath with multi-hash computation in a single kernel, and targets multiple kernel instantiations on a single FPGA with nonce-based \ac{HBM} partitioning. We verified the computation modules under simulation; however, its memory interface demands improvements for random access rather than the simulator's straightforward memory models. 

Our CryptoNight-Haven miner RTL, testbench, and host-code (a software patch to XMRig \cite{xmrig} with \ac{XRT} integration) are publicly available at: \url{https://github.com/KULeuven-COSIC/CryptoNightHaven-FPGA-miner}.

\section{Algorithm}

The Cryptonight-Haven algorithm is a variant of Cryptonight \cite{Cryptonote}. It chains together multiple well-known cryptographic primitives: AES, Keccak, Blake, etc. Collectively, these primitives are used to initialize a large scratchpad data buffer, on which semi-random operations are performed.

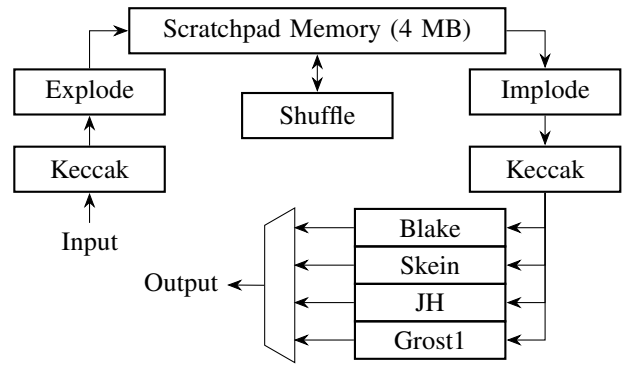
\begin{figure}[t]
\centering
\input{figures/overview2.tex}
\caption{The dataflow Cryptonight-Haven. 
}
\label{fig:overview}
\end{figure}

\Cref{fig:overview} shows an overview of the computation flow. First, the input passes through a Keccak module, extracting the state. The explode module takes the first 32 bytes of the state and expands them to 10 AES round keys. These keys are used to perform AES rounds on the remaining bytes (divided into 8 blocks of 128-bits) of the Keccak state. After 10 rounds, the AES output is written to the memory buffer. Next, the blocks are XOR'ed with each other, and they again undergo 10 rounds of AES with the same keys as before. This process is repeated until 4 MB of data has been generated.

The Shuffle step performs a semi-random operation; either AES, a division, or a multiplication followed by addition. Corresponding input and output data yield intensive memory accesses to irregular addresses.

The Implode operation is similar to Explode. Bytes 32-63 of the state are used to generate AES round keys. Bytes are read from the start of the memory buffer and XOR'ed with state bytes 64-191. Subsequently, these bytes are put through 10 AES rounds --one round per key-- and the next bytes in the memory buffer are read. After reading the entire 4 MB scratchpad buffer twice, the 10 AES rounds are repeated 16 additional times without reading from the memory buffer.

Finally, the new state passes through a Keccak permutation to generate the final state. Depending on the 2 LSBs of this state, it is subsequently hashed using either of the Blake, Skein, JH or Groestl schemes, resulting in the final 256-bit output.

\section{Implementation}

Our single accelerator kernel employs a datapath with a module hierarchy similar to \Cref{fig:overview}. The hash computations in the first and last steps of the computation stages contribute a negligible overhead. The Explode step consists of a simple implementation that generates 4 MB of data with simple binary operations and stores it into \ac{HBM} memory. It heavily benefits from AXI burst transfers with multiple outstanding transactions. The Implode step has double the latency of Explode for accessing the memory twice. For the underlying computations, both the Explode and Implode modules employ 10 AES cores.

In contrast, Shuffle's computation is demanding due to the multitude of iterations and underlying memory accesses. Additionally, data dependencies between consecutive accesses prevent optimizations, e.g. burst transfers. The memory size prevents using on-chip memory --BRAM or URAM-- that allows single-cycle data access. Irregular addresses are another obstacle that prevents caching parts of the memory on-chip. Hence, minimizing data transfer overheads is the most advantageous strategy.

The memory accesses of Shuffle form one of the foundations for CryptoNight-Haven's ASIC-resistance claims. Implementing these computations on an ASIC requires significant chip resources to be reserved for implementing memory, leaving a limited amount of silicon for the computation. In contrast, the Varium C1100 natively has 8~GB of \ac{HBM} available, and our accelerator heavily benefits from it. Moreover, the partitioning of \ac{HBM} into a number of \acp{PC} allows us to instantiate each Shuffle unit with a dedicated memory port, resulting in a scalable design.

We pipelined Shuffle for simultaneous computation of up to 128 hashes to boost the computation performance. That requires an identical increase in memory consumption --easily accommodated by the 8~GB HBM-- where individual nonces for each hash computation partition memory regions. In line with this pipelining, we split the computation into various stages that communicate over AXI-Stream interfaces connected with FIFOs. Our pipelining approach allows the time-critical Shuffle module to be clocked at 500~MHz while the other modules remain at 200~MHz.

\section{Future Work}

In its current state, our design computes hashes correctly within the Vivado simulation environment. That employs a simplified view of memory, which restricts the memory model to AXI accessed BRAM. However, when instantiated as an \ac{XRT} kernel on the Varium C1100, the hash computations are inconsistent with simulation. We have enhanced the design with a set of AXI-lite accessible status registers that collect additional performance and debug information on the hardware execution. The further roadmap we envision is as follows:

\begin{enumerate}
    \item Enhancing our Vivado simulation with random AXI access latencies and AXI protocol checkers.
    \item Progressing with \ac{XRT} kernel construction, by replacing the BRAM with HBM under Vitis \texttt{hw\_emu} based \ac{XRT} executions.
    \item Extending Shuffle's memory accesses with Xilinx' \ac{RAMA} IP.
\end{enumerate}

After these steps enable the correct computation of the CryptoNight-Haven \ac{PoW}, we have already taken the first steps to integrate the accelerator into XMRig \cite{xmrig} using \ac{XRT} APIs. The accelerator should be compared thoroughly to existing CPU and GPU-based miners for Cryptonight-Haven, hopefully showing increased throughput and/or energy efficiency. Finally, we also aim to compare to related work: FPGA-based miners were proposed for the ASIC-resistant \ac{PoW} Lyra2REv2~\cite{Lyra2-FPGA, Lyra2-standalone}, Scrypt \cite{MRSA21}, and X16R \cite{9786081}.

\bibliographystyle{IEEEtran}
\bibliography{references}

\end{document}

%% file: figures/tikzconfig.tex
\tikzset{
    include sha/.code={
    \def\shawidth{15mm}
    \def\shasep{3mm}
    },
    line cap = rect,
         port/.style={font=\tiny},
         module/.style={font=\ttfamily, rectangle, thin, draw=black, minimum width = 6mm, minimum height = 6mm},
         main/.style={rectangle, thick, draw=black, fill=red!30, minimum width = 10mm, minimum height = 14mm},
         controledge/.style={dashed, dash pattern = on 1pt off 1pt, blue},
         dataedge/.style={black, thick},
         mux/.style={trapezium, text width = #1, minimum height = 3mm, trapezium left angle=120, trapezium right angle = 120, thin, draw=black},
         sha3/.style={module, minimum width=\shawidth, minimum height=6mm},
         hv path/.style = {to path={-| (\tikztotarget)}},
         vh path/.style = {to path={|- (\tikztotarget)}},
         s path/.style = {to path={-- ++(0,#1) -| (\tikztotarget)}},
         v path/.style = {to path={-- ++(#1,0) |- (\tikztotarget)}},
         register/.style = {rectangle, thin, draw=black, minimum width = 6mm, minimum height = 10mm, append after command = {
             \pgfextra
                 \draw[line cap = rect] (\tikzlastnode.north west) ++ (\pgflinewidth, -1.5mm) -- ++ (0.5mm, -0.5mm) -- ++ (-0.5mm, -0.5mm);
             \endpgfextra}
         }
}

%% file: figures/AES_construction.tex
\pgfkeys{/tikz/.cd,
    /tikz/ampersand replacement=\&,
    at/.initial={(0,0)},
    at/.get=\coordpos,
    at/.store in=\coordpos,
    AES block/.code={
        \matrix (#1 center) [column sep = 2mm, row sep = 2mm] at \coordpos {
            \node (#1 imux) [mux=1mm,rotate=90] {}; \&
            \node (#1 block) [register] {};\& 
            \node (#1 AES) [module] {AES};\\
            \& \node (#1 key) [register] {};\&\\
        };
        \graph [use existing nodes] {
            #1 imux -- #1 block -- #1 AES;
            #1 key --[hv path] #1 AES.south;
        };
        \coordinate (#1 muxup) at ($(#1 imux.north)+(0,1.5mm)$);
        \coordinate (#1 muxdown) at ($(#1 imux.north)+(0,-1.5mm)$);
    }
}

%% file: figures/overview2.tex
\begin{tikzpicture}[]

  \tikzset{
    comp/.style   = {draw, thick, minimum width=2.00cm,  minimum height=0.6cm },
    memory/.style = {draw, thick, minimum width=5.00cm,  minimum height=0.6cm },
    hash/.style   = {draw, thick, minimum width=2.00cm,  minimum height=0.5cm },
    sarrow/.style = {-{Stealth[scale=1.2]}},
    darrow/.style = {{Stealth[scale=1.2]}-{Stealth[scale=1.2]}},
    mux/.style    = {trapezium, 
      trapezium left angle  = 120, 
      trapezium right angle = 120, 
      text width = 0.7cm, 
      rotate=270, 
      minimum height = 4mm, 
      draw=black,
      xshift=0.925cm,
      yshift=0.00cm}
  }

  \node[comp]   (E)  []                  {Explode};
  \node[comp]   (S)  [right=1.00cm of E.south east] {Shuffle};
  \node[comp]   (I)  [right=1.00cm of S.north east] {Implode};
  \node[memory] (M)  [above=0.50cm of S] {Scratchpad Memory (4~MB)};
  \node[comp]   (Ki) [below=0.40cm of E] {Keccak};
  \node[comp]   (Ko) [below=0.40cm of I] {Keccak};
  
  \node[hash]   (HB) [below left = 0.20cm and -0.5cm of Ko] {Blake};
  \node[hash]   (HS) [below      =-0.03cm            of HB] {Skein};
  \node[hash]   (HJ) [below      =-0.03cm            of HS] {JH};
  \node[hash]   (HG) [below      =-0.03cm            of HJ] {Grost1};
  
  \node[mux]    (MX) [left=2cm of HS.south] {};
  
  \node[]       (ID) [below=0.40cm of Ki      ] {Input};
  \node[]       (OD) [left =2.70cm of HS.south] {Output};
  
  \draw [sarrow] (ID.north) -- (Ki.south);
  \draw [sarrow] (Ki.north) -- (E.south);
  \draw [sarrow] (E.north)  |- (M.west);
  \draw [darrow] (S.north)  -- (M.south);
  \draw [sarrow] (M.east)   -| (I.north);
  \draw [sarrow] (I.south)  -- (Ko.north);
  \draw [sarrow] (Ko.south)  |- (HB.east);
  \draw [sarrow] (Ko.south)  |- (HS.east);
  \draw [sarrow] (Ko.south)  |- (HJ.east);
  \draw [sarrow] (Ko.south)  |- (HG.east);
  
  \draw [sarrow] (HB.west) -- ++ (-0.8cm, 0);
  \draw [sarrow] (HS.west) -- ++ (-0.8cm, 0);
  \draw [sarrow] (HJ.west) -- ++ (-0.8cm, 0);
  \draw [sarrow] (HG.west) -- ++ (-0.8cm, 0);
  
  \draw [sarrow] (MX.south) -- (OD.east);

\end{tikzpicture}